# Liquid Metal as Connecting or Functional Recovery Channel for the Transected Sciatic Nerve

Jie Zhang [1], Lei Sheng [1], Chao Jin [1], and Jing Liu [1, 2*]

1. Department of Biomedical Engineering, School of Medicine, Tsinghua University,
Beijing 100084, China

2. Technical Institute of Physics and Chemistry, Chinese Academy of Sciences,
Beijing 100190, China

**\*Address for correspondence:**

Dr. Jing Liu

Department of Biomedical Engineering,

School of Medicine,

Tsinghua University,

Beijing 100084, P. R. China

E-mail address: jliubme@tsinghua.edu.cn

Tel. +86-10-62794896

Fax: +86-10-82543767

**Abstract:**

In this article, the liquid metal GaInSn alloy (67% Ga, 20.5% In, and 12.5% Sn by volume) is proposed for the first time to repair the peripheral neurotmesis as connecting or functional recovery channel. Such material owns a group of unique merits in many aspects, such as favorable fluidity, super compliance, high electrical conductivity, which are rather beneficial for conducting the excited signal of nerve during the regeneration process in vivo. It was found that the measured electroneurographic signal from the transected bullfrog's sciatic nerve reconnected by the liquid metal after the electrical stimulation was close to that from the intact sciatic nerve. The control experiments through replacement of GaInSn with the conventionally used Riger's Solution revealed that Riger's Solution could not be competitive with the liquid metal in the performance as functional recovery channel. In addition, through evaluation of the basic electrical property, the material GaInSn works more suitable for the conduction of the weak electroneurographic signal as its impedance was several orders lower than that of the well-known Riger's Solution. Further, the visibility under the plain radiograph of such material revealed the high convenience in performing secondary surgery. This new generation nerve connecting material is expected to be important for the functional recovery during regeneration of the injured peripheral nerve and the optimization of neurosurgery in the near future.

## 1. Introduction

In modern society, peripheral nerve injury is an increasing source of morbidity worldwide, which often results in restricted activity or life-long disability.[1] Despite the progress in the cognition on anatomy and pathophysiology of the peripheral nerve system, as well as innovations in microsurgical techniques and materials, the management of peripheral nerve injuries remains a big challenge so far.[2] The repair and regeneration of the peripheral neurotmesis caused by accidental trauma, disease, or surgical procedures has been found rather tough, which usually requires extremely complex surgical intervention, lengthy process of regeneration and may have certain probability of functional disability after regeneration.

So far, various microsurgical treatments and materials have been investigated for optimal repairing and regeneration of the transected peripheral nerve. The method of direct suturing of the cut ends (nerve stumps) is commonly employed when the gap is no longer larger than 20mm, while if a significant nerve gap exists, the necessity of nerve graft, nerve transfers or tubulization techniques stand out.[3] Since the infection of the nerve stumps or the retraction due to their elasticity often increase the extent of the gap, the reconnection and regeneration of transected peripheral nerve with large defects turns out a core issue in neurosurgery. As the current gold standard, nerve autografting[4] becomes the most common way when the direct suturing is not preferred. However, disadvantages of nerve autografting, such as the need for a secondary surgery, significant donor site morbidity and limited length of available graft resource make it a problematical method. In contrast to nerve autografting, nerve allografting makes it possible to bridge larger nerve defects, [5] but the requirement of systemic immunosuppression lasts for 18 months. And during this time, patients are prone to opportunistic infections and tumor formation. [6] The reconstruction with the method of nerve transfers [7] depends on the donor nerve close to the transected peripheral nerve. In comparison with auotologous nerve grafts, the demand for long nerve grafts by way of nerve transfers is avoided, while it may result in disfunction of the donor nerve muscle. In order to find out an alternative method, various natural and synthetic biomaterials have been investigated as nerve conduits in experimental and clinical conditions. [8] As only several commercially available synthetic nerve conduits have been approved by the U.S. FDA, [9] and the therapeutic effect of such conduits appears only feasible for the short gap case, both the materials and fabrication technique require tremendous efforts in tubular implants and scaffolds.

Apart from the methods of bridging the transected peripheral nerve, acceleration of peripheral nerve regeneration has been extensively researched. [10] Growth factors, support cells or insoluble extracellular matrices have been demonstrated to promote peripheral nerve regeneration. Although various strategies of microsurgical treatments and materials for peripheral neurotmesis have been enormously improved over the past decades, the outcome in repairing the peripheral neurotmesis remains unsatisfactory as the full functional recovery is seldom achieved. [11] In addition, peripheral nerve regeneration is a lengthy process. Taken axonal regeneration as an example, the speed of regeneration and restoration approximates 1 mm/d. [12] Years may be needed for reinnervation because of the large nerve defect. Over this period, degeneration will occur, and the denervated end organs undergo histologic changes that are consistent with muscle atrophy and infiltration of fibrous tissue. Consequently, target muscle atrophy becomes irreversible, which may lead to the impaired physiological function, resulting in long-term disabilities. [13] Therefore, taking the functional recovery into consideration during the regeneration is of great significance for patients.

Recently, the eutectic GaInSn alloy has been found to own several intrinsic properties, which may be advantageous as peripheral nerve functional recovery channel during the regeneration of transected peripheral nerve. With a broad temperature range of liquid phase, the room temperature liquid metal GaInSn alloy (67% Ga, 20.5% In, and 12.5% Sn by volume) may make it more convenient in fabrication, encapsulation and surgical operation. [14] Theoretically, such liquid metal is zero stiffness and almost infinite stretchability due to its fluidity, [15] which is super compliant in vivo. The electrical conductivity of such material is $3.1*10^6$ $Sm^{-1}$, which is several orders higher than those of non-metallic materials and in contrast with many other metallic materials commonly used in electric transmission.[16] Therefore, it may minimize ohm loss and conduct weak electroneurographic signal maximally. The chemical compatibility, tunable wettability, and the ability to be directly printed on a wide variety of materials [17] such as polymers, glass, and metals make it cooperate with nerve conduits harmoniously.

In the present work, the liquid metal GaInSn was proposed for the first time to treat peripheral neurotmesis as functional recovery channel during the regeneration of the nerve. The animal study has been approved by the Ethics Committee of Tsinghua University, Beijing, China under contract [SYXK (Jing) 2009-0022]. As the first trial in this direction, the liquid metal used in the experiments was a kind of eutectic of Ga, In and Sn alloy with a volume proportion of 67%, 20.5% and 12.5%, respectively. The detailed synthesis procedures were outlined in former study.[18] We demonstrated the feasibility and advantages of such material through the reconnection of the transected bullfrog sciatic nerve in vitro. To evaluate the basic electrical property of GaInSn alloy, its resistivity and impedance were measured. Besides, the visibility under the plain radiograph was investigated. Different from the conventional neurosurgery, the innovative material employed in the repair of peripheral nerve is expected to efficiently conduct electroneurographic signal during the regeneration, and avoid degeneration or even target muscle atrophy.

**2. Materials and Methods**

The experimental subject was established as the sciatic nerve with gastrocnemius from bullfrogs under the consideration that it has been serving as one of the most typical specimens in the neurological test and easily available. Ten living bullfrogs were bought from the local market. The gastrocnemius with sciatic nerve was extracted with the double pithing method. The Riger's Solution was prepared and applied to infiltrate the sciatic nerve and gastrocnemius in order to keep them active. After preparation of the experimental specimen, the intact sciatic nerve with gastrocnemius was infiltrated in the Riger's Solution for twenty minutes. Since the electrophysiological stimulation is one of the dispensable methods in neuroscience research to induce synaptic plasticity and to probe synaptic function,[19] the voltage signal was employed to stimulate the proximal nerve.

The experimental system to measure the electroneurographic signal was shown in **Figure 1**. Two glass dishes filled up with Riger's Solution were used to contain the proximal sciatic nerve and gastrocnemius with distal sciatic nerve, and the glass slide was applied to position the nerve and isolate Riger's Solution. Two methods to encapsulate the liquid metal or Riger's Solution were shown in **Figure 1** (**b**). One was that a piece of transparent plastic with a rectangular groove used for loading liquid metal or Riger's Solution to reconnect the transected sciatic nerve was placed on the glass slide. The size of the groove was only 10mm*1mm*0.5mm, small enough to reduce the dosage of the liquid metal. Another was that the liquid metal or Riger's Solution was filled in a glass capillary pipette with the length of 3cm and diameter of 0.5mm. The digital signal generator (Agilent 33210A, Agilent Technologies, U.S.A.) has been used to produce the electric-stimuli signal (Wave form: square-wave, Voltage amplitude: 0.2V, Frequency: 0.5Hz, Duty cycle: 20%). The data acquisition device (Agilent 34970A, Agilent Technologies, U.S.A.) with the resolution of 0.001mV was employed to measure the electroneurographic signal after the simulation. The electroneurographic signal was displayed and collected by a computer. With the simulation of the sciatic nerve, the innervated gastrocnemius became contracted. In order to record the dynamic behavior of the nerve subject to stimulation, the high-speed camera Canon XF305 was adopted.

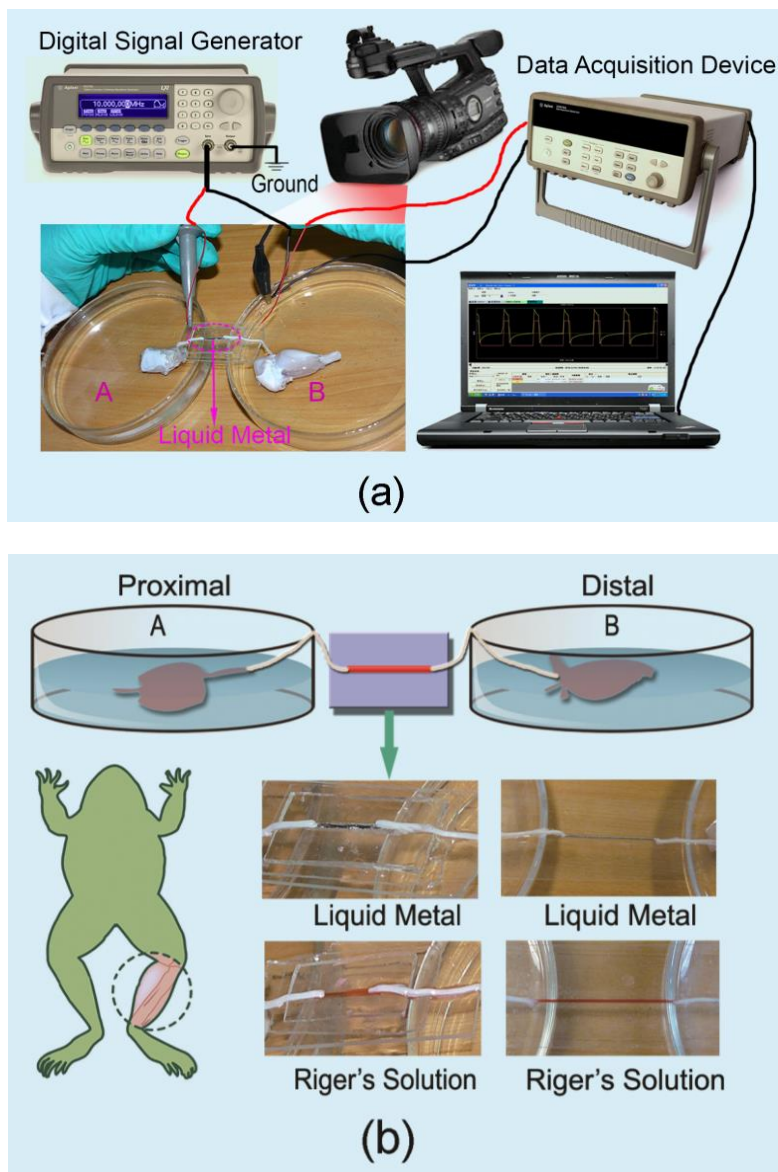

**Figure 1**. The experimental diagram for the overall system. (a) The experimental setup of measuring the electroneurographic signals. (b) The schematic diagram of the transected sciatic nerve reconnected by liquid metal and Riger's Solution, respectively. To view it clearly, a little red ink was added into the Riger's Solution.

The negative terminals of both digital signal generator and data acquisition device were linked with the glass dish containing the distal nerve with gastrocnemius together. The positive terminal of the stimulation signal was loaded on the proximal nerve, while the positive terminal of the data acquisition device was put on the distal nerve, about 2cm away from the stimulation signal, as shown in **Figure 1** (**a**). Before the sciatic nerve was transected, the electroneurographic signal of the intact nerve after stimulation was collected by the data acquisition device. The behavior of the gastrocnemius was recorded by the high-speed camera simultaneously. Then the nerve was transected totally by ophthalmic scissors. In order to test the performance of functional reconnection, the transected nerve was reconnected by the liquid metal or Riger's Solution and the electroneurographic signal as well as dynamic contraction after the same stimulation was recorded.

In order to compare the electrical properties of the liquid metal GaInSn with the Riger's Solution and characterize the electrical property quantitatively, the micro-ohmmeter (Agilent 34420A, Agilent Technologies, U.S.A.) was employed. The resistance of the liquid metal or Riger's Solution encapsulated in a glass capillary pipette with the length of 10cm and diameter of 0.5mm was measured under the direct voltage. Besides, the SRS Model SR780 2-Channel Network Signal Analyzer was adopted to measure the impedance of the liquid metal or Riger' Solution encapsulated in a plastic tube with the length of 40cm and diameter of 1mm from 1Hz to 1kHz, and each group was measured five times to obtain the average results.

It is expected the liquid metal could cooperate with biodegradable nerve conduits during the regeneration. Therefore, after the regeneration, if it is no longer necessary for keeping the liquid metal, the simple secondary surgery may be required. The visibility under the plain radiographs and fluidity could make it convenient to suck up the liquid metal via a micro-syringe. To investigate the visibility under the plain radiographs, the liquid metal was injected into the bullfrog's leg where it was close to the sciatic nerve, and then the plain radiograph was acquired by X-ray under the dosage of 100V and 1mAs. After that, the liquid metal was sucked up by a micro-syringe, and the plain radiograph was acquired once again.

## 3. Results and Discussion
### *3.1. The electroneurographic signal*

A carefully selected electric-stimuli signal (Wave form: square-wave, Voltage amplitude: 0.2V, Frequency: 0.5Hz, Duty cycle: 20%) was applied to stimulate the proximal nerve. As the electric-stimuli signal could excite the nerve, the measured waveform of electroneurographic signal on the intact nerve was not the same as the original electric-stimuli signal, which contained a positive phase wave and a negative phase wave during every periodicity, as shown in **Figure 2** (**a**) and (**b**). The waveform was basically synchronized with the electric-stimuli signal and similar to the electric-stimuli signal in some degree. After the sciatic nerve was transected and reconnected through the selected liquid metal or Riger's Solution, the electroneurographic signal on the distal nerve was collected (**Figure 2** (**a**) and (**b**)). Though the electroneurographic signal from the transected sciatic nerve was virtually similar to that from the intact sciatic nerve, they were subtly different at the strategic locations. In **Figure 2** (**b**), the difference was quite obvious, mainly at the peaks and the troughs. The statistics of the amplitudes of the electroneurographic signals and the changes at the peaks and troughs were shown in **Figure 2** (**c**). The changes of the voltage on the transected distal nerve reconnected by the liquid metal were more coincided with that reconnected by Riger's Solution. In addition, the same performance was obtained when the liquid metal or Riger's Solution was encapsulated in a glass capillary pipette with the length of 3cm and diameter of 0.5mm, respectively. By stimulating the intact sciatic nerve or reconnecting the transected sciatic nerve via liquid metal or Riger's Solution, the gastrocnemius innervated by the sciatic nerve was contracted (see supplementary materials).

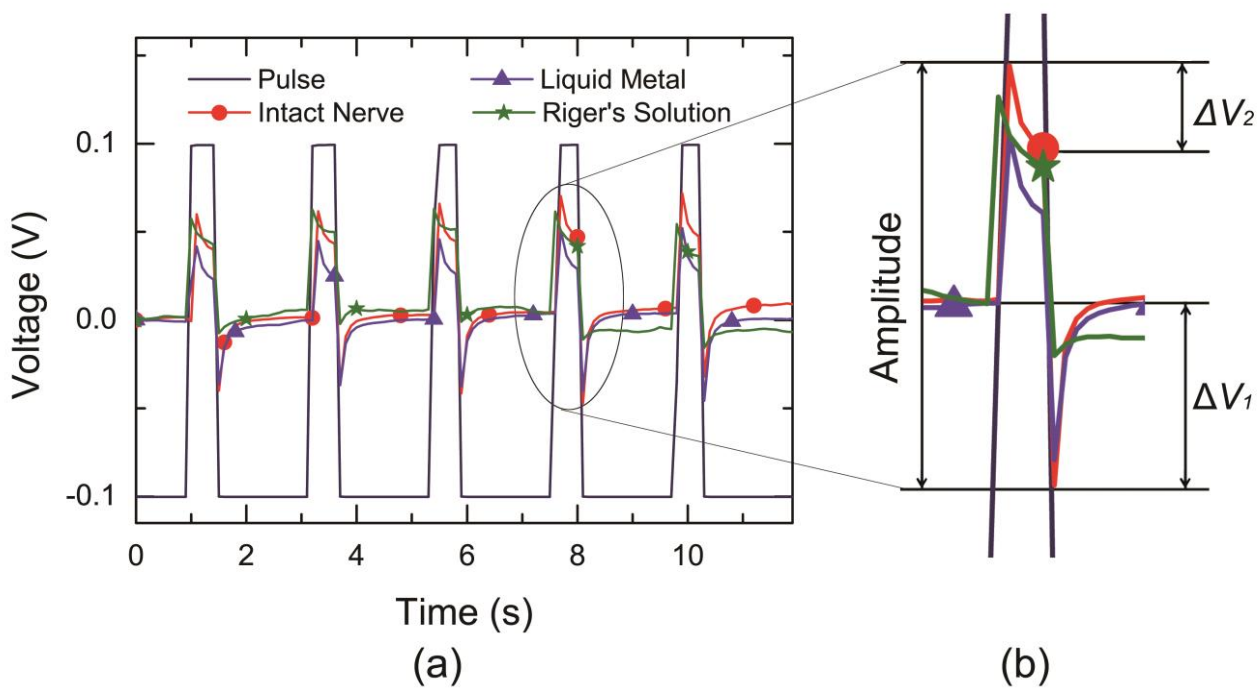

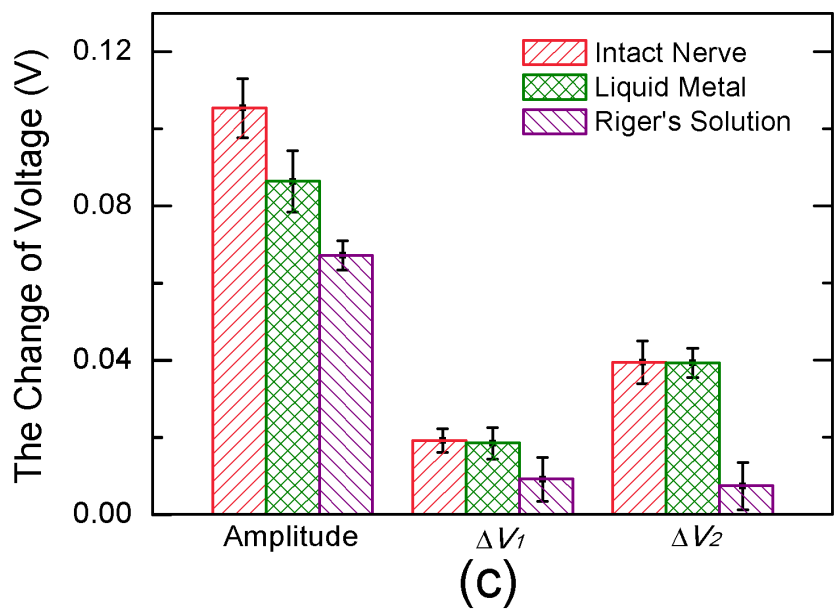

**Figure 2**. The electroneurographic signal. (a) The electric-stimuli signal and the excitement signal from the intact nerve, the transected nerve reconnected by GaInSn or Riger's Solution with the volume of 10mm*1mm*0.5mm, respectively. (b) Amplified detail of partial view (a). (c) The comparison of voltage changes after electrical stimulation.

As the sciatic nerve is composed by various kinds of nerve fibers, the recorded electroneurographic signals were compound nerve action potentials. In the neural electrophysiology, after the nerve receives effective stimulus, the action potentials could be generated. The action potentials on the nerve trunk were the compound action potentials from a number of nerve fibers. Thus, the characteristic of the action potentials on the nerve trunk was different from that on the single nerve fiber: the waveform was biphasic, which contained a positive phase and a negative phase. As seen in **Figure 2**, if removing the $\Delta V_1$ and $\Delta V_2$, the rest of the electroneurographic signals might mainly reflect the electric leakage by Riger's Solution, which was inevitable since the existence of the Riger's Solution was necessary to avoid the neurological impairment during the measurement. The real part that could denote the compound nerve action potentials was $\Delta V_1$ and $\Delta V_2$. The absolute value $\Delta V_2$ was larger than $\Delta V_1$, as shown in **Figure 2** (**c**), mainly because the electric leakage might cover up the positive phase. The electric leakage was more obvious when the transected sciatic nerve was reconnected through the Riger's Solution due to complete contact with the extracellular matrix. The applied liquid metal GaInSn is insoluble. Therefore, when the liquid metal was encapsulated in the nerve conduits, it could stay in the conduits and perform the conduction function of the electroneurographic signals stably.

*3.2. The comparison of physical properties*

The resistance of the used liquid metal and Riger's Solution was 0.0224 Ω, 0.443787 MΩ, respectively (measured at room temperature 20 $^o$C in a glass capillary pipette with the length of 10mm and diameter of 0.5mm under direct current). Accordingly, the calculated resistivity of the liquid metal used and Riger's Solution was $4.396*10^{-8}$ Ω m and 0.8709 Ω m, respectively, as shown in **Table 1**. The resistivity of the Riger's Solution was several orders higher than the liquid metal GaInSn, which might result in the attenuation of electroneurographic signals. Hence, Riger's Solution was not suitable for conducting the weak electroneurographic signals. The melting point of the liquid metal was 10.5 $^o$C, which indicated that it remained in the liquid phase over the body temperature (37 $^o$C).

**Table 1.** Comparison of physical properties of GaInSn with Riger's Solution

|  | Density (kg m$^{-3}$) | Melting point ($^o$C) | Resistivity (Ω m) |
| --- | --- | --- | --- |
| GaInSn | 6360 | 10.5 | $4.396*10^{-8}$ |
| Riger's Solution | 1009 | <4 | 0.8709 |

From the view of impedance, both the resistance and the reactance of the applied GaInSn alloy were several orders lower than that of Riger's Solution, as shown in **Figure 3**. Since the frequencies of most electroneurographic signals are below 10 kHz, the influence of the GaInSn reactance can be ignorable and hence the resistance remains the major determinant of the impedance. In addition, the resistance of the GaInSn alloy was relatively stable from 1 Hz to 10 kHz. While for Riger's Solution, the resistance and reactance were highly affected by frequency, which indicated that Riger's Solution was not stable if it was applied for conducting dynamic signals.

Furthermore, the viscosity of the applied GaInSn is $2.98*10^{-7}$ $m^2/s$,[15] which elucidated that such material was super compliant in body and would not influence the mechanical properties of the nerve conduits. In general, the basic physical properties indicated that the GaInSn owns unique advantages for conducting electroneurographic signals in vivo.

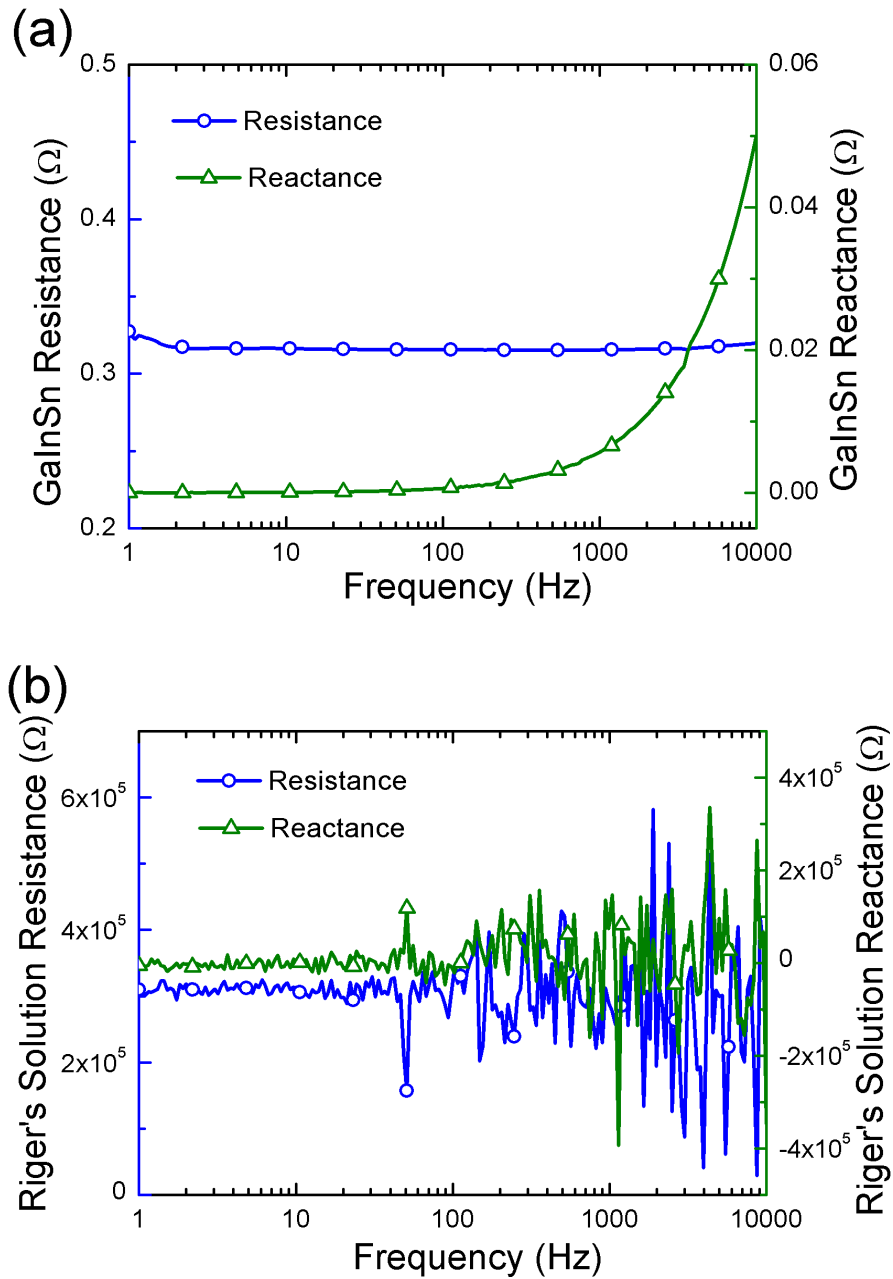

**Figure 3**. The impedance of the applied liquid metal GaInSn alloy and Riger's Solution encapsulated in a plastic tube with length of 40cm and diameter of 1 mm. (a) The resistance and reactance curve of the liquid metal. (b) The resistance and reactance curve of Riger's Solution.

## 3.3. The visibility under plain radiograph

Under the plain radiograph, such material showed favorable visibility, as shown in **Figure 4** (**b**). The liquid metal is close to the femur, as well as the sciatic nerve (seen in **Figure 4** (**a**) and (**b**)). If liquid metal cooperates with the biodegradable nerve conduits, the visibility may contribute to the surgical navigation. Under the guidance of the plain radiograph, the liquid metal can be sucked up by a micro-syringe when the regeneration completed, and the complex secondary surgery is thus avoided.

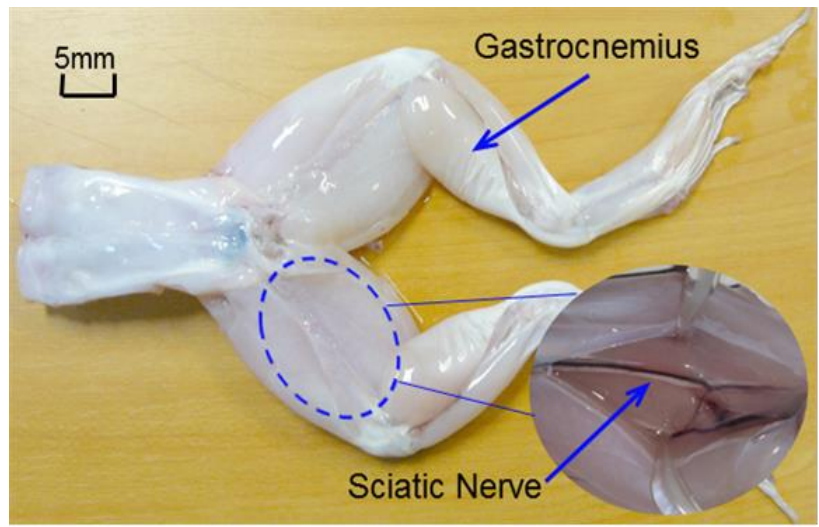

(a)

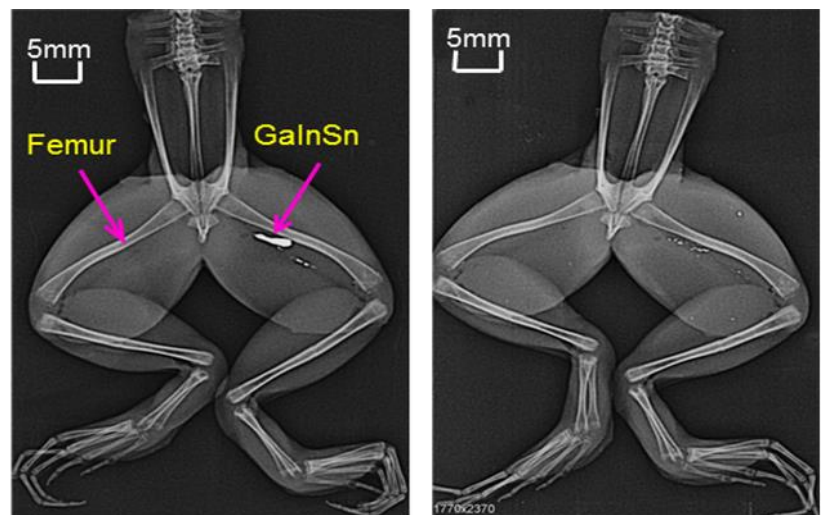

(b)　　　　　　　　　　(c)

**Figure 4**. The pictures of bullfrog's lower body. (a) The photograph of bullfrog's lower body after injecting GaInSn alloy. (b) The plain radiograph of bullfrog's lower body after injecting GaInSn alloy. (c) The plain radiograph of bullfrog's lower body after sucking up GaInSn alloy with a micro-syringe.

It is necessary to note that the reason to choose Riger's Solution as the control group is that it matches well with biological tissue without causing any safety problem. Though favorable compatibility and safety it owns, Riger's Solution is not suitable as the functional recovery channel due to its high resistivity and assimilation with the tissue. In contrast to Riger's Solution, the applied liquid metal is insoluble and not absorbed. Concerns may be mentioned about the safety of the liquid metal in body. Fortunately, GaInSn is a chemically benign metal alloy.[20] Though the accurate electrochemical corrosion of such material needs further study, the main ingredient gallium has already been applied in medicine in many aspects.[21]

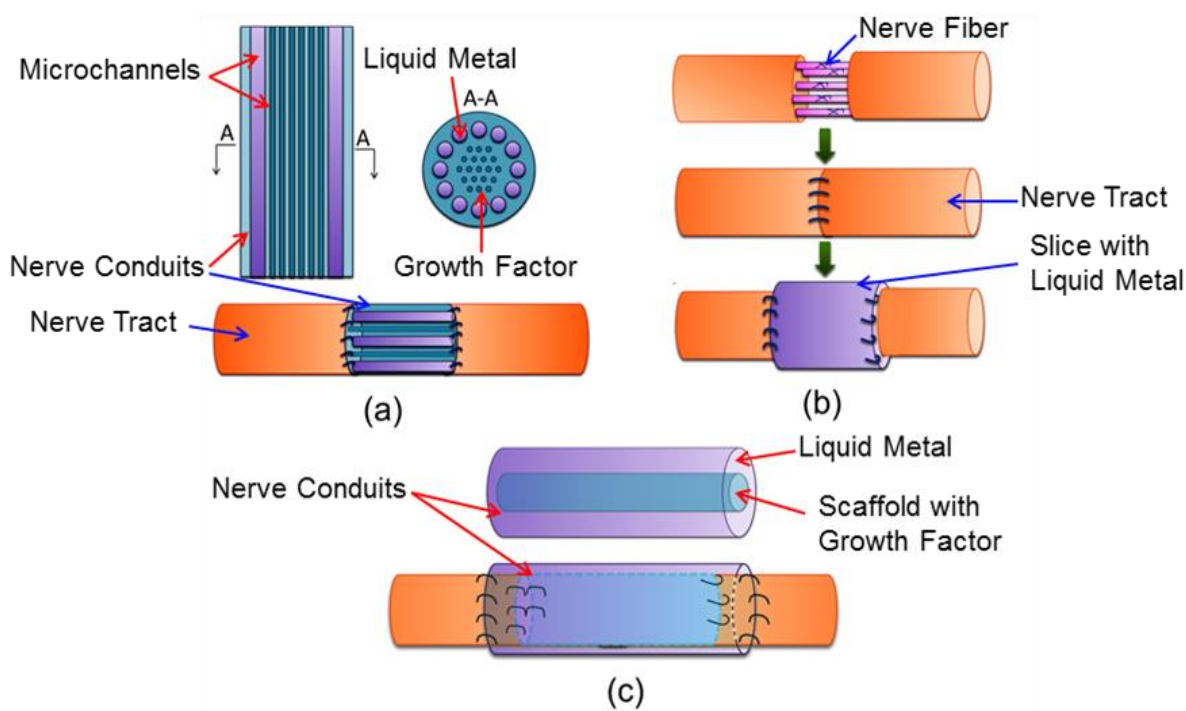

**Figure 5**. Three kinds of nerve conduits to repair the injured peripheral nerve. (a) Nerve conduit with microchannels. (b) Nerve conduit with a shape of thin slice. (c) Nerve conduit with concentric tubes.

The liquid metal GaInSn shows great virtues that no other conventional material possesses, and it is promising to achieve functional recovery during the regeneration if combining the nerve conduits with liquid metal. It is of great concern for patients to alleviate the sufferings from peripheral neurotmesis and reduce the risk of physiological function impairment or even long-term disabilities. The related fabrication and encapsulation may not be complex, as many fashioned nerve conduits can be employed with small changes. One of the tentative ideas is to take advantage of the conventional nerve conduits with microchannels,[22] and fill the peripheral thicker channels with the liquid metal, as shown in **Figure 5** (**a**). The paracentral thinner microchannels are filled with materials like growth factors, Schwann cell or insoluble extracellular matrices to promote the regeneration. Another tentative idea is to fabricate the nerve conduits with concentric tubes, in which the internal cylinder is filled with porous support materials and substances to promote the regeneration, while the periphery is filled with the liquid metal to conduct the electroneurographic signals, as shown in **Figure 5** (**c**). Apart from peripheral neurotmesis, it is expected that neurological decline due to disease or accidental crash can be improved by wrapping a biocompatible slice coating with liquid metal to enhance the conduction of electroneurographic signals, as shown in **Figure 5** (**b**).

**4. Conclusion**

In summary, this work performed the first ever experiment to demonstrate that liquid metal GaInSn alloy could efficiently reconnect the transected sciatic nerve in vitro and conduct the electroneurographic signals. The applied liquid metal GaInSn was presented as an alternative medium over existing ones and might play an important role of the functional recovery channel cooperating with nerve conduits during regeneration. The unique diverse favorable properties of the GaInSn guarantee it an advantageous material for future clinical practices. Besides, the applied liquid metal could represent with similar electrical properties of the nerve. It is expected that such kind of materials could open a promising way for nerve functional recovery during the regeneration process and reduce the probability of functional impairment.

**Acknowledgements:**

This work was partially supported by NSFC under Grant # 51376102.

**References**

[1] a) A. D. Widgerow, A. A. Salibian, E. Kohan, T. Sartiniferreira, H. Afzel, T. Tham, G. R. Evans, *Microsurgery*. **2013**, 1; b) J. Noble, C.A. Munro, V. S. Prasad, R. Midha, *J. Trauma*. **1998**, *45*, 116.

[2] J. Scheib, A. Höke, Nat. Rev. Neurol. **2013**, *9*, 668

[3] I. V. Yannas, M. Zhang, M. H. Spilker, *J. Biomater. Sci., Polym. Ed.* **2007**, *18,* 943.

[4] a) H. Millesi, *Clin. Plast. Surg.* **1984**, *11*, 105; b) H. Millesi, *Surg. Clin. North. Am.* **1981**, *61*, 321.

[5] G. R. Evans, *Anat. Rec.* **2001**, *1*, 396.

[6] A. Pabari, S. Y. Yang, A. M. Seifalian, A. Mosahebi, *J. Plast. Reconstr. Aesthet. Surg*. **2010**, *63*, 1941.

[7] A. M. Moore, C. B. Novak, *J. Hand Ther.* **2014**, 1.

[8] a) X. Jiang, S. H. Lim, H. Mao, S. Y. Chew, *Exp. Neurol.* **2010**, *223*, 86; b) X. Zhan, M. Gao, Y. Jiang, W. Zhang, W. M. Wong, Q. Yuan, H. Su, X. Kang, X. Dai, W. Zhang, J. Guo, W. Wu, *Nanomedicine* **2013**, *9*, 305; c) S. Madduri, K. Feldmanb, T. Tervoort, M. Papaloïzos, B. Gander, *J. Controlled Release* **2010**, *143*,168; d) T. W. Chung, M. C. Yang, C. C. Tseng, S. H. Sheu, S. S.Wang, Y. Y. Huang, S. D. Chen, *Biomaterials* **2011**, *32*, 734; e) A. Pabari, S. Y. Yang, A. Mosahebi, A. M. Seifalian, *J. Controlled Release* **2011**, *156*,2.

[9] S. Kehoe, X. F. Zhang, D. Boyd, *Injury* **2012**, *43*, 553.

[10] a) M. Ikeda, T. Uemura, K. Takamatsu, M. Okada, K. Kazuki, Y. Tabata, Y. Ikada, H. Nakamura, *J. Biomed. Mater. Res., Part A*, **2013**, 266; b) C. H. E. Ma, T. Omura, E. J. Cobos, A. Latrémolière, N. Ghasemlou, G. J. Brenner, E. van Veen, L. Barrett, T. Sawada, F. Gao, G. Coppola, F. Gertler, M. Costigan, D. Geschwind, C. J. Woolf. *J. Clin. Invest.* **2011**, *121*, 4332.

[11] a) A. M. Hart, G. Terenghi, M. Wiberg, *Neurol. Res.* **2008**, *30*, 999; b) M. Wiberg, G. Terenghi, *Surg. Technol. Int.* **2003**, *11*, 303.

[12] M. G. Burnett, E. L. Zager, *Neurosurg. Focus* **2004**, *16*, 1.

[13] J. E. Shea, J. W. Garlick, M. E. Salama, S. D. Mendenhall, L. A. Moran, J. P. Agarwal, *J. Surg. Res.* 2013, 1.

[14] C. Jin, J. Zhang, X. Li, X. Yang, J. Li, J. Liu, *Sci. Rep.* **2013**, *3*, 3442.

[15] N. B. Morley, J. Burris, L. C. Cadwallader, M. D. Nornberg, *Rev. Sci. Instrum.* **2008**, *79*, 056107.

[16] Y. Liu, M. Gao, S. Mei, Y. Han, J. Liu, *Appl. Phys. Lett*. **2013**, *102*, 064101.

[17] Y. Zheng, Z. He, Y. Gao, J. Liu, *Sci. Rep.* **2013**, *3*, 1786.

[18] Y. Gao, H. Li, J. Liu, *PLoS One* **2012**, *7*, 1.

[19] W. S. Anderson, F. A. Lenz, *Nat. Clin. Pract. Neurol*. **2006**, *2*, 310.

[20] F. M. Blair, J. M. Whitworth, J. F. McCabe, *Dent. Mater.* **1995**, *11*, 277.

[21] a) S. M. Dunne, R. Abraham, *Br. Dent. J.* **2000**, *189*, 310; b) W. C. Chen, K. D. Tsai, C. H. Chen, M. S. Lin, C. M. Chen, C. M. Shih, W. Chen, *Intern. Emerg. Med.* **2012**, *7*, 53; c) F. B. Hagemeister, S. M. Fesus, L. M. Lamki, T. P. Haynie,1990. *Cancer* **1990**, *65*, 1090.

[22] a) S. P. Lacoura, R. Attab, J. J. FitzGeraldc, M. Blamired, E. Tarteb, J. Fawcettc, *Sens. Actuators A* **2008**, *147*, 456; b) T. Cui, Y. Yan, R. Zhang, L. Liu, W. Xu, X. Wang, *Tissue Eng. Part C* **2009**, *15*, 1.